\documentclass[a4paper]{aa}
\usepackage{apj2aa}
\usepackage{astron} 
\usepackage{float,epsfig,psfig}
\usepackage{pstricks}
\def\hea4{{\it HEAO~A4}}
\def\heaoa2{{\it HEAO~A2}}
\def\heao1{{\it HEAO~1}}

\def\eg{{\it e.g.}~}

\def\h0{$H_{\rm o}=50$~km~s$^{-1}$~Mpc$^{-1}$}
\def\q0{$q_{\rm o}$}

\def\msun     {M$_{\odot}$}

\def\etal    {{et~al.}~}

\def\cms3  {~{cm$^{-3}$}}

\begin{document}

\title{X-ray Evidence for Spectroscopic Diversity of Type Ia Supernovae:}
\subtitle{XMM observation of the elemental abundance pattern in M87.} 
\author{Alexis Finoguenov\inst{1}, Kyoko Matsushita\inst{1},  Hans
  B\"ohringer\inst{1}, Yasushi Ikebe\inst{1}
\and 
Monique Arnaud\inst{2}}


\offprints{A. Finoguenov, alexis@xray.mpe.mpg.de}

\institute{Max-Planck-Institut f\"ur extraterrestrische Physik,
             Giessenbachstra\ss e, 85748 Garching, Germany 
\and
 Service d'Astrophysique, CEA Saclay, 91191 Gif sur Yvette Cedex, France}

\date{Received July 4 2001; accepted October 22 2001}
\authorrunning{Finoguenov \etal}
\titlerunning{Spectroscopic Diversity of Type Ia Supernovae}

\abstract{
We present the results of a detailed element abundance study of hot gas in
M87, observed by XMM-Newton. We choose two radial bins, $1'-3'$ and $8'-16'$
($8'-14'$ for EMOS; hereafter the central and the outer zones), where the
temperature is almost constant, to carry out the detailed abundance
measurements of O, Ne, Mg, Si, S, Ar, Ca, Fe and Ni using EPIC-PN (EPN) and
-MOS (EMOS) data. First, we find that the element abundance pattern in the
central compared to the outer zone in M87 is characterized by SN Ia
enrichment of a high (roughly solar) ratio of Si-group elements (Si, S, Ar,
Ca) to Fe, implying that Si burning in SN Ia is highly incomplete. In
nucleosynthesis modeling this is associated with either a lower density of
the deflagration-detonation transition and/or lower C/O and/or lower central
ignition density and observationally detected as optically subluminous SNe
Ia in early-type galaxies. Second, we find that SN Ia enrichment has a
systematically lower ratio of the Si-group elements to Fe by 0.2 dex in the
outer zone associated with the ICM of the Virgo cluster.  We find that such
a ratio and even lower values by another 0.1 dex are a characteristic of the
ICM in many clusters using observed Si:S:Fe ratios as found with ASCA.
Third, the Ni/Fe ratio in the central zone of M87 is $1.5\pm0.3$ solar
(meteoritic), while values around 3 times solar are reported for other
clusters. In modeling of SN Ia, this implies a reduced influence of fast
deflagration SN Ia models in the chemical enrichment of M87's ISM.  Thus, to
describe the SN Ia metal enrichment in clusters, both deflagration as well
as delayed detonation scenarios are required, supporting a similar
conclusion, derived from optical studies on SNe Ia. We discuss implications
of our results for observations of SN Ia and also suggest a link between the
element abundances observed in the M87 X-ray halo with the observed
characteristics of M87 globular clusters.  
\keywords{galaxies: individual:
M87 --- galaxies: clusters: individual: Virgo --- X-rays: galaxies ---
supernovae: general }
}

\maketitle
\section{Introduction}

Element abundance patterns favoring a strong role of SNe Ia in the iron
group production, as indicated by solar and subsolar ratios of alpha-process
elements (Mg and Si) to iron, were found in elliptical galaxies, groups of
galaxies, and centers of clusters of galaxies (Finoguenov \etal 1999;
Finoguenov \& Ponman 1999; Hwang \etal 1999; Finoguenov, David, Ponman 2000,
hereafter FDP; Dupke \& White 2000b). As a consequence of the strong role of
SN Ia in the enrichment of the hot gas, X-ray observations are well suited
to put constraints on the element production by SN Ia.

Compared to previous results, largely provided by ASCA, the high collecting
area of the XMM X-ray telescopes and the sensitivity at low and high
energies now offers the possibility also to study the O, Ar and Ca
abundance, as well as to improve on the statistical significance of
measurements of other elements, like Si, S, Fe, Ni. Addition of the O
abundance is particularly important for estimation of the SN II
contribution, since other indicators, like Ne and Mg are flawed due to their
proximity to iron L-shell complex (Mushotzky \etal 1996). Although, some O
abundance estimates were already available with ASCA, calibration problems
impose systematic uncertainties in the obtained results (\eg\ Mushotzky
\etal 1996), so ASCA's estimates of the SN II role were based on the Si/Fe
ratio, which has the disadvantage of Si being also contributed by SN
Ia. Therefore knowledge of the Oxygen abundance provides the best means to
separate the enrichment by the two types of supernovae.

\begin{table*}
{
\begin{center}
\footnotesize
{\renewcommand{\arraystretch}{1.3}
\caption{\footnotesize
XMM EPIC-PN and EPIC-MOS results on element abundances in M87.
\label{tab:ab}}

\begin{tabular}{rcccccc}
\hline
\multicolumn{1}{c}{Z, A}  & \multicolumn{2}{c}{EPN} & \multicolumn{2}{c}{EMOS}\\
 & $Z_c$ $(1'-3')$ & $Z_o$ $(8'-16')$ & $Z_c$ $(1'-3')$ & $Z_o$ $(8'-14')$ & units$^{\natural}$  & theory$^{\sharp}$\\
\hline
 O 16.00 & 0.546 (0.521:0.566) &0.366 (0.341:0.402) &0.535 (0.514:0.554) &0.386 (0.365:0.411) & 8.51e-4 & 8.51e-4\\
Si 28.11 & 1.121 (1.075:1.145) &0.602 (0.557:0.650) &1.242 (1.223:1.260) &0.657 (0.632:0.673) & 3.55e-5 & 3.58e-5\\
S  32.10 & 0.933 (0.884:0.964) &0.406 (0.360:0.457) &0.973 (0.951:0.995) &0.406 (0.379:0.434) & 1.62e-5 & 1.84e-5\\
Ar 36.33 & 0.953 (0.843:1.040) &0.463 (0.322:0.587) &0.851 (0.793:0.905) &0.392 (0.318:0.465) & 3.63e-6 & 3.62e-6\\
Ca 40.13 & 1.429 (1.273:1.562) &0.786 (0.597:0.975) &1.296 (1.207:1.379) &0.622 (0.515:0.718) & 2.29e-6 & 2.19e-6\\
Fe 55.92 & 0.599 (0.577:0.624) &0.336 (0.308:0.363) &0.650 (0.632:0.668) &0.344 (0.320:0.360) & 4.68e-5 & 3.23e-5\\
Ni 58.78 & 1.337 (1.163:1.628) &                    &2.573 (1.655:3.497) &0.800 (0.000:2.532) & 1.78e-6 & 1.77e-6\\
\hline            
\end{tabular}
}     
\end{center}

{\footnotesize {\sc Note:} --- errors are quoted on the 68\% confidence
level. All element abundances are derived using the {\it K-shell} lines.}

%
%

\begin{center}
\footnotesize
{\renewcommand{\arraystretch}{1.3}
\caption{\footnotesize
Element abundances, dependent on the modeling of Fe L-shell complex.
\label{tab:ab2}}

\begin{tabular}{rcccccc}
\hline
\multicolumn{1}{c}{Z, A}  & \multicolumn{2}{c}{EPN} & \multicolumn{2}{c}{EMOS}\\
 & $Z_c$ $(1'-3')$ & $Z_o$ $(8'-16')$ & $Z_c$ $(1'-3')$ & $Z_o$ $(8'-14')$ & units$^{\natural}$ & theory$^{\sharp}$ \\
\hline
Ne 20.15 &0.370 (0.271:0.483) &0.447 (0.334:0.563) &0.419 (0.356:0.485) &0.290 (0.215:0.353) & 1.23e-4 & 1.24e-4\\
Mg 24.34 &0.590 (0.542:0.633) &0.208 (0.145:0.277) &0.630 (0.595:0.659) &0.238 (0.203:0.275) & 3.80e-5 & 3.84e-5\\
Ni 58.78 &0.963 (0.852:1.078) &0.612 (0.444:0.789) &0.886 (0.800:0.965) &0.978 (0.550:1.068) & 1.78e-6 & 1.77e-6\\
\hline            
\end{tabular}
}     
\end{center}

{\footnotesize {\sc Note:} --- errors
are quoted on the 68\% confidence level. The Ni abundance is derived using the
Ni {\it L-shell} lines.}

\vspace*{0.4cm}

$^{\natural}$ \hspace*{0.3cm}{\footnotesize Reference abundance values, used
in our (and most X-ray) abundance determinations with the results tabulated
in columns (2--5). These units correspond to solar photospheric values from Anders
\& Grevess (1989).}

$^{\sharp}$ \hspace*{0.3cm}{\footnotesize Solar abundances, as tabulated in
Woosley and Weaver (1995). These units correspond to meteoritic values from
Anders \& Grevess (1989) and are used in Figs.1-4 to simplify the comparison
with the commonly used logarithmic values of the abundance, $[Z/H]$.}

}
\end{table*}

With addition of observational data on many extra elements, X-ray
observations become sensitive to various ad-hoc assumptions in the modeling
of SNe explosion. For example, the O/Si ratio in SN II ejecta can place
constraints on the energy release during the explosion (Nakamura \etal
2000). In this regard, the observation of M87, where SN Ia products dominate
in the abundance pattern, potentially leads to a refinement of SN Ia
models. Since enrichment of the intracluster gas typically requires
$10^{9-10}$ SN Ia ($10^{8}$ on scales of a BCG), it presents an extremely
good averaging of any possible difference in the scenarios for SN Ia
explosions. Since SNe Ia are adopted as standard candles for high-redshift
distance measurements, understanding of SN Ia enrichment in clusters, where
accumulation of SN Ia products might have started at redshifts of about 2
(Finoguenov, Arnaud, David 2001), can help to minimize the systematics of
these measurements, resulting most importantly in constraints on the value
for $\Lambda$ (H\"oflich, Wheeler \& Thielemann 1998).

Data on SN Ia reveal that brighter events tend to have a broader
light-curve. This behavior is caused by a variation in a SN Ia yield for
$^{56}Ni$ (e.g. Mazzali \etal 2001) and is modeled by a variation in the
density of the deflagration-detonation transition (DDT) (Khokhlov 1991,
however see Niemeyer 1999). Allowed densities typically take a range of
$0.5-3\times10^7$ g cm$^{-3}$ and are reached (with time) by the expanding
WD from originally a few times $10^9$ g cm$^{-3}$ (therefore these models
bear the name delayed detonation). A lower transition density is associated
with a reduced $^{56}$Ni production and consequently with a dimmer SN Ia.
Umeda \etal (1999) proposed that it is the mass fraction of Carbon in the CO
WD that controls the DDT, such that a lower Carbon mass fraction corresponds
to a dimmer event. In addition, Umeda \etal (1999) point out that a smaller
C/O ratio produces a small buoyancy force, thus leading to a slower deflation
and to a dimmer SN. Using C/O as a determining factor, Umeda \etal (1999)
showed that the observed systematics in SN Ia could be consistently
explained from the point of view of the galactic evolution, as the C/O ratio
varies systematically with stellar mass (Umeda \etal 1999a). Some SN Ia may
not exhibit any DDT, but instead an acceleration of the burning front
propagation from initially a few per cent up to 30 per cent of the sound
speed.  These models are called convective deflagration (Nomoto \etal 1984,
Iwamoto \etal 1999). In our modeling we will use the nucleosynthesis
calculations of Nomoto \etal (1997), where convective deflagration is
represented by the W7 model and three delay-detonation models WDD1, WDD2,
and WDD3 are given, where the last digit refers to the DDT density in units
of $10^7$ g cm$^{-3}$.  Differences in nucleosynthesis yields among the
models, important for our observation, could be subdivided into two groups:
Si-Ca/Fe and Ni/Fe ratios. The former is an indication of incompleteness of
the Si group burning into the iron group and the latter, which is a tracer of
$^{58}$Ni/$^{56}$Fe is an indicator of neutron-rich isotope production,
which depends on the efficiency of electron capture in the core of the
exploding white dwarf. Fast deflagration, like the original W7 model of
Nomoto \etal (1984) and its combinations with slow initial deflagration
(WS15W7), typically yields larger $^{58}$Ni/$^{56}$Fe ratio, compared to
slow initial deflagration followed by detonation: $^{58}$Ni/$^{56}$Fe is
4.3, 2.7, 1.5, 1.3 solar for W7, WS15W7, WS15DD3, WS15DD1 models,
respectively (Iwamoto \etal 1999).

The particular interest in the observation of M87 comes from its proximity,
which allows us to resolve the temperature structure, complications due to
X-ray emission of central AGN, the jet and the radio-lobes and to achieve
high photon statistics, needed in order to determine less abundant
elements. In view of the finite spatial resolution of the X-ray telescopes,
we can derive element abundances much closer to the center in M87 than for
any other central cluster region and thus measure the element abundance
patterns within a region mostly dominated by SN Ia ejecta.

\section{Results}\label{s:res}

This {\it Paper} reports a detailed analysis of the element abundances from
the data of the XMM-Newton (Jansen \etal 2001) performance verification
observation of M87, described elsewhere (B\"ohringer \etal 2001; Belsole
\etal 2001). In the following analysis we will exclude the spatial zones
contaminated by the X-ray emission from the jet and radio-lobes. Initial
results of the determination of element abundances in M87 are presented in
B\"ohringer \etal (2001) and Molendi \& Gastaldello (2001), where strong
gradients where found in the Fe, Si and S abundance, while the O abundance
did not show significant variations. In the central half arcminute, an
abundance decrease in all the elements is observed, which may be explained
by resonance scattering and temperature structure (Matsushita \etal 2001a).
The de-projected temperature profile exhibits two shelves at $0.5'-3'$ and
$8'-16'$ radii (hereafter the central and the outer zones), at the level of
1.6 and 2.5 keV (for more details of the de-projection technique and the
choice of the regions for spectrum extraction see Matsushita \etal 2001b,
hereafter MBFB). Therefore, at these radii, while achieving high statistics
in the spectral extraction we still have a simple temperature model, which
assures the robustness of our results. For the central region, the removal
of projection effects of the outer cluster emission is done by subtracting
the scaled spectra from outer parts. Since the cluster gas temperature is
nearly isothermal for radii far beyond $16'$, such a subtraction is not
necessary for the outer zone. We use the XMMSAS version of April 2001, where
\eg\ for EPN it is for the first time possible to estimate and remove the
induced background due to the Out-of-Time Events, include split events into
the spectral analysis, and to produce a refined CTI (charge transfer
inefficiency) correction. We use the vignetting corrected data according to
the in-flight calibrations. We use Lockman hole observations for background
subtraction. The EMOS and EPN detector response matrices used are from the
June 2001 release. More details on the primary data reduction are given in
MBFB.

For a 17 Mpc distance to M87, the angular scale is 5 kpc/arcminute and the
radial boundaries for the inner and the outer zones correspond to 5--15 kpc
and 40--80 kpc, respectively. We note that for our choice of the extraction
radii, the effects of resonance scattering are not important. For example
the optical depth for O lines responsible for the abundance determination is
less than 0.4 (Matsushita \etal 2001a). While such an estimate is
complicated for Fe, we note that the Fe abundances derived using L- and
K-shell lines agree well with each other, while effects of resonance
scattering differ.

Two specific approaches applied in the spectral modeling have to be noted:
to account for the calibration uncertainties in the low-energy response of
the EPN we limit the spectral fits to energies above 0.5 keV. To avoid
problems with the mismatch between the low and high energy vignetting in the
EMOS calibration files we fit the low and high energy part of the EMOS
spectrum for the outer zone separately. The residuals in the EMOS spectral
fitting in the 1.2--2.0 keV energy range, which result in a 10\% higher Si
abundance compared to the EPN values, are attributed to calibration problems
and cannot be corrected for at present. This also reduces the confidence in
the Al abundance determination. Therefore we postpone the study of the Al
abundance to a later work.

Fits to the EPN data for the outer zone are carried out in the 0.5--7 keV
band, because at higher energies the spectrum is dominated by the
background. This prevents us from a determination of the Ni abundance with
the EPN using K-shell lines for the outer zone. In all other cases, the
spectral fitting is done using the 0.5--10 keV energy band. In determining
the data for Table \ref{tab:ab}, where only measurements using K-shell lines
are presented, we excise the 0.7-1.6 keV energy range from spectral
analysis, which also removes a dependence of the derived temperature on the
shape of Fe L-shell peak. The temperature is determined instead by the shape
of the bremsstrahlung radiation at high energies. We use the MEKAL model
(Mewe \etal 1985, Mewe and Kaastra 1995, Liedahl \etal 1995), fixing the He,
C and N abundance to their solar values.  Given the systematics in using the
present-day plasma codes to fit the Fe L-shell line complex (Phillips \etal
1999), some care is needed in the abundance determination using K-shell
lines for Ne and Mg (Mushotzky \etal 1996) and L-shell lines for Ni, as
their lines are located within the blend of Fe L-shell lines. The exact
spectral shape of this blend is sensitive to the presence of a low
temperature component, identified in the central part of M87 by MBFB. 

\includegraphics[width=8.4cm]{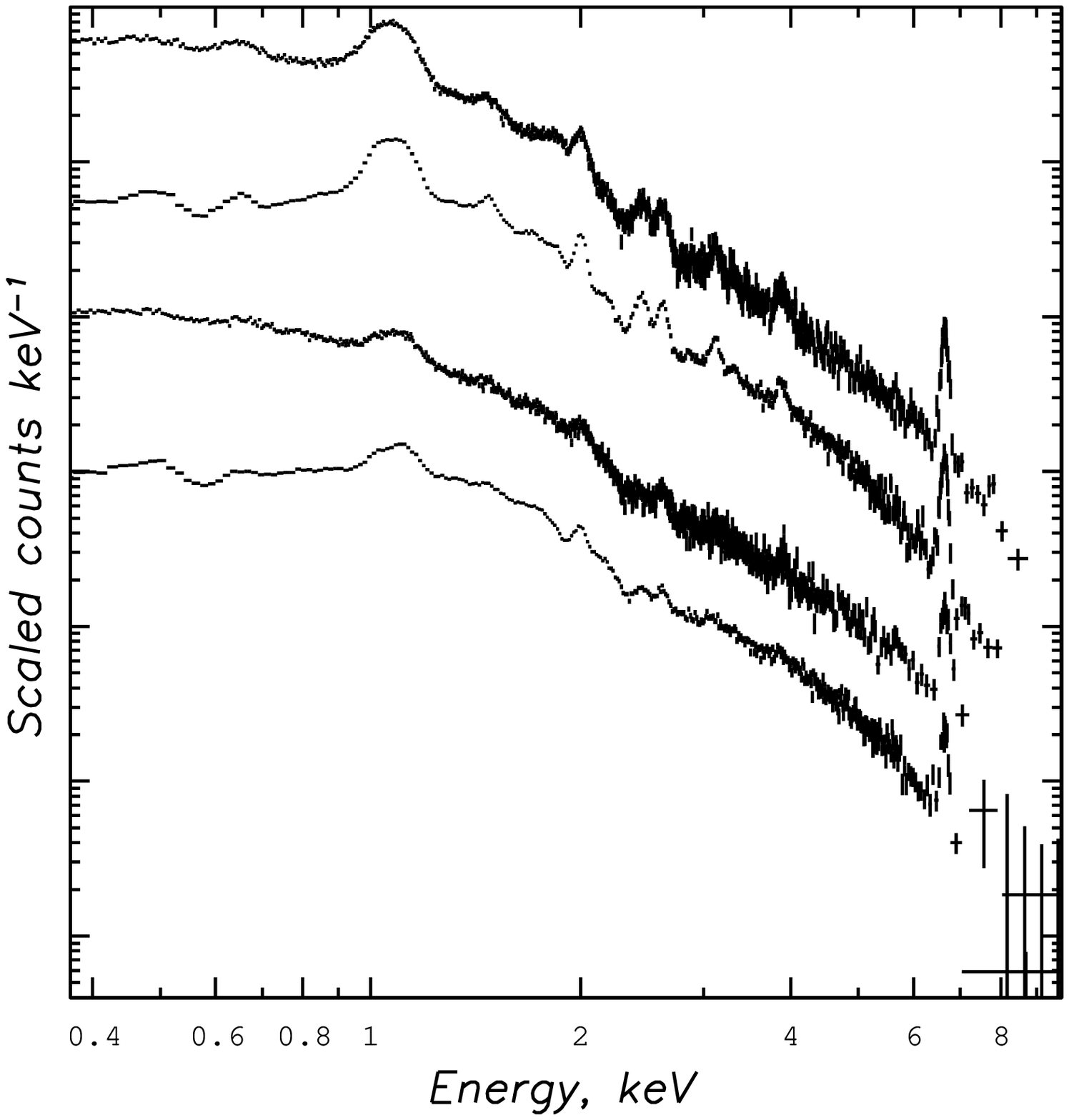}

\figcaption{XMM spectra used to derive the heavy element abundance of M87.
From top to bottom: EPN and EMOS spectra for the central zone, EPN and EMOS
spectra for the outer zone.
\label{fig-spe}
}

We also observe the residuals in the spectral fitting due to uncertainties
in the MEKAL plasma code and uncertainties in the detector energy
response. To check the influence of residual temperature structure, we
performed the spectral extraction in $0.5'-2'$, $1'-2'$ and $1'-3'$ annuli,
also ignoring projection effects. Comparison of the determined element
ratios to Fe shows remarkably similar results, while residuals in spectral
fits to Fe L-shell lines are smallest for $1'-2'$ and $1'-3'$. We also
conclude that just using the extracted spectra from $1'-3'$ (excluding the
jet) without performing the de-projection, results in exactly the same
abundance ratios. Since we find the projection effects to be negligible, we
present the results of the simple spectral extraction of data in the $1'-3'$
annulus. This allows us to increase the statistics of the measurement,
important for abundance measurements of rare elements and makes it easier
for other observers to reproduce our results. Spectra, used in the final
analysis are displayed in Fig.\ref{fig-spe}. For the determination of the Fe
abundance as listed in Table \ref{tab:ab} we use K-shell lines, while to
model the spectral presence of Fe L-shell lines we add an additional
spectral component with separately fitted temperature and verify the results
by applying the {\it APEC} code (Smith \etal 2001), available in XSPEC v11,
in place of the MEKAL code. As the confidence in the abundance determination
for Ne, Mg and Ni in this way is still somewhat compromised, we present our
results for them separately (Table \ref{tab:ab2}). A solar Mg/O abundance
ratio is detected for the central zone, while in the outer zone the Mg/O
abundance ratio is slightly sub-solar. In the outer zone, however, when we
fix the temperature and the iron abundance on the level determined by the
high energy continuum and the Fe K-shell lines, solar Mg/O is reproduced for
EPN, while such analysis of EMOS data is complicated, as we cannot use the
normalization of the continuum derived at high energies, due to insufficient
calibration of the vignetting below $\sim2$ keV, as discussed above and in
MBFB.  We therefore conclude that the Mg/O ratio is consistent with the
solar value for the outer zone in M87.

\includegraphics[width=8.4cm]{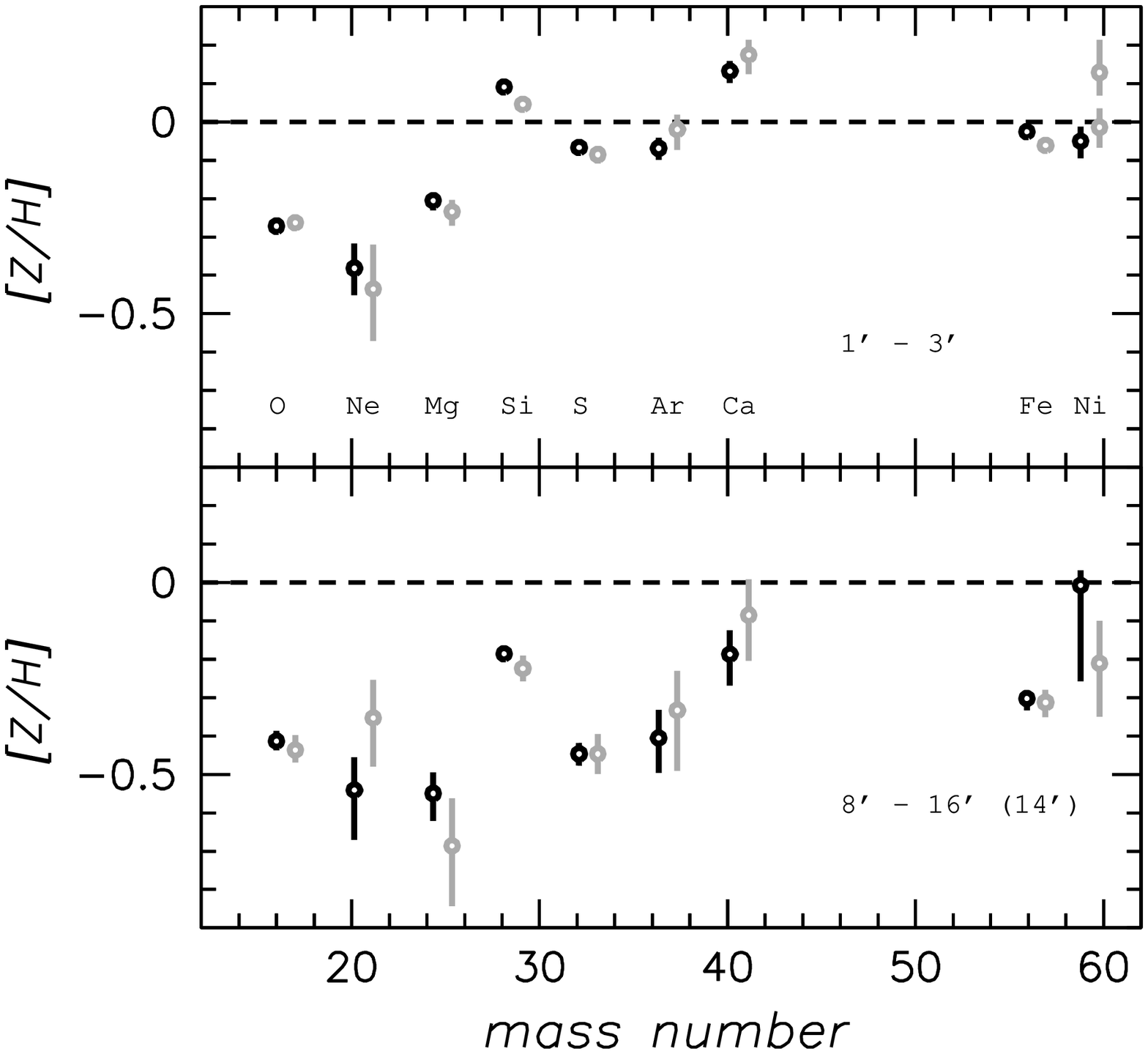}

\figcaption{Abundance of heavy elements in M87 derived in the central
($1'-3'$, upper panel) and outer ($8'-16'$ EPN and $8'-14'$ EMOS, lower
panel) zones. EMOS data are shown in black and EPN data in grey. EPN data
points are shifted by 1 in mass number for clarity of the comparison.  Error
bars are shown at the 68\% confidence level. Abundances are presented in the
logarithmic units relative to the solar meteoritic values, as adopted in
theory of chemical enrichment and listed in Table \ref{tab:ab}. Dashed line
on both panels indicate the solar abundance. The data on Ni abundance
correspond to the measurements using Ni L-shell lines. In addition we show
the EPN measurement of Ni abundance in the central zone using the K-shell
line (highest of the two grey points on Ni in upper panel), which has
comparable error bars.
\label{ab-data}
}

\begin{figure*}
 
 \hfill\includegraphics[width=7.cm]{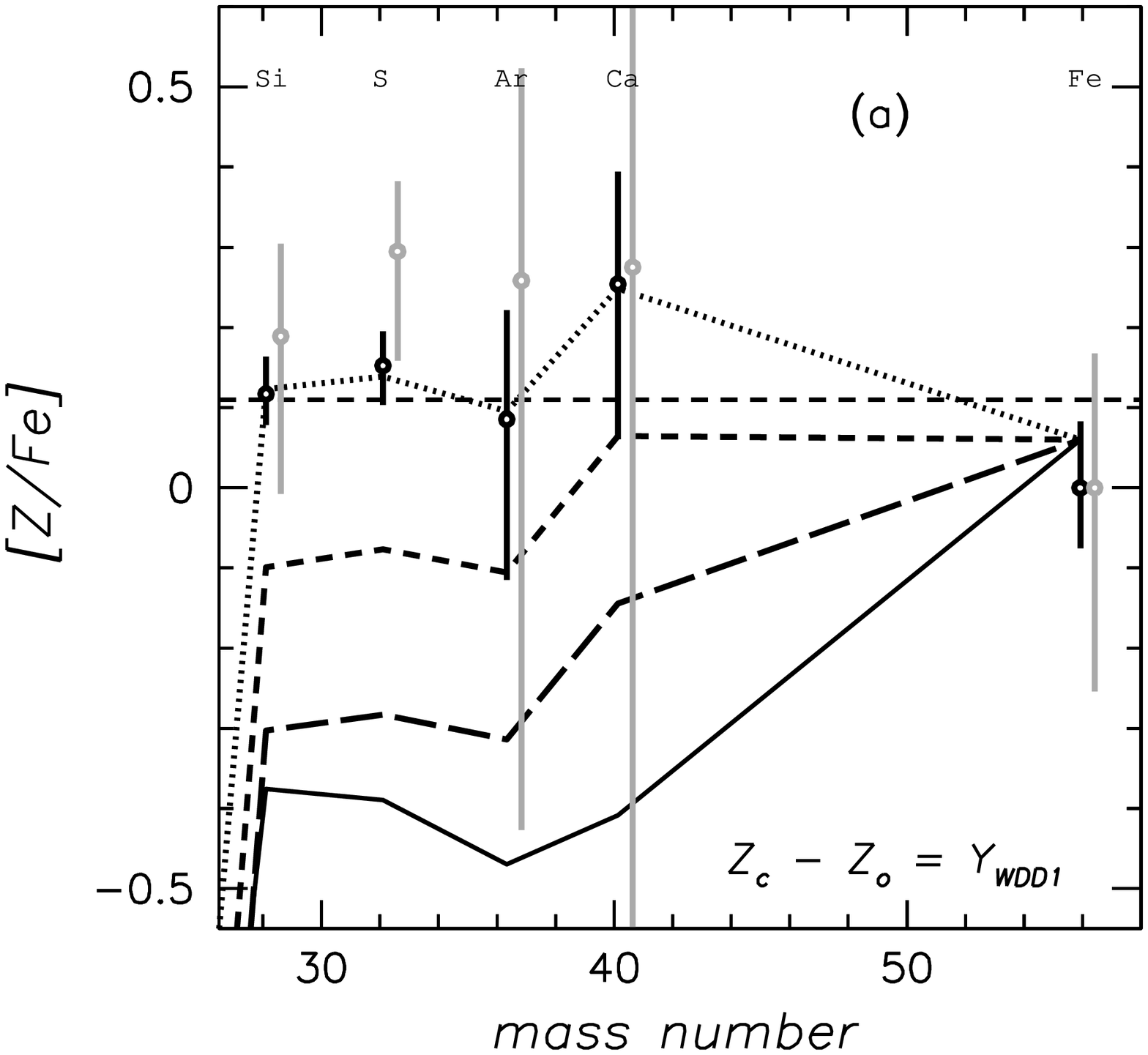}\hfill\includegraphics[width=7.cm]{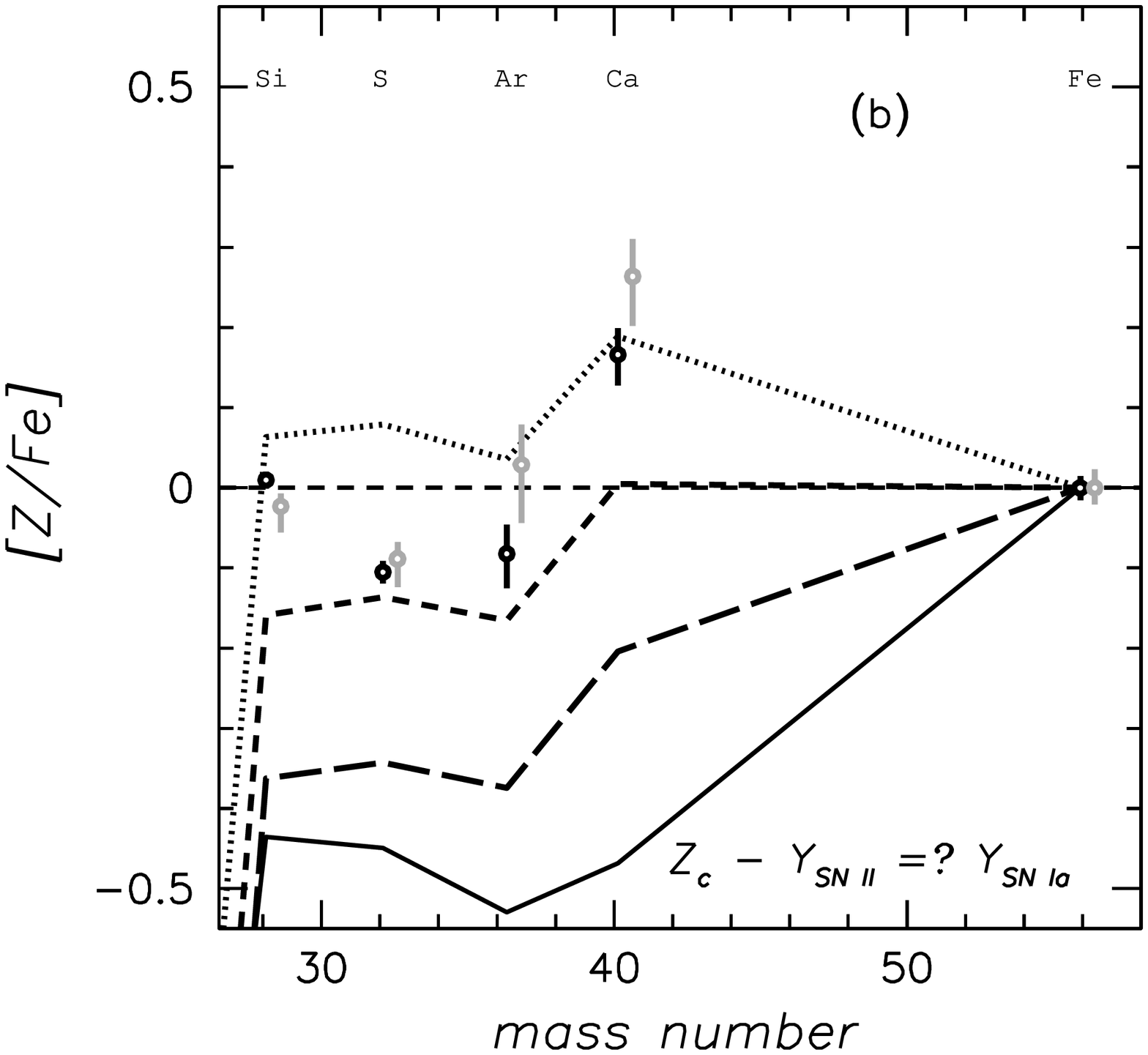}\hfill\hfill
 
 \hfill\includegraphics[width=7.cm]{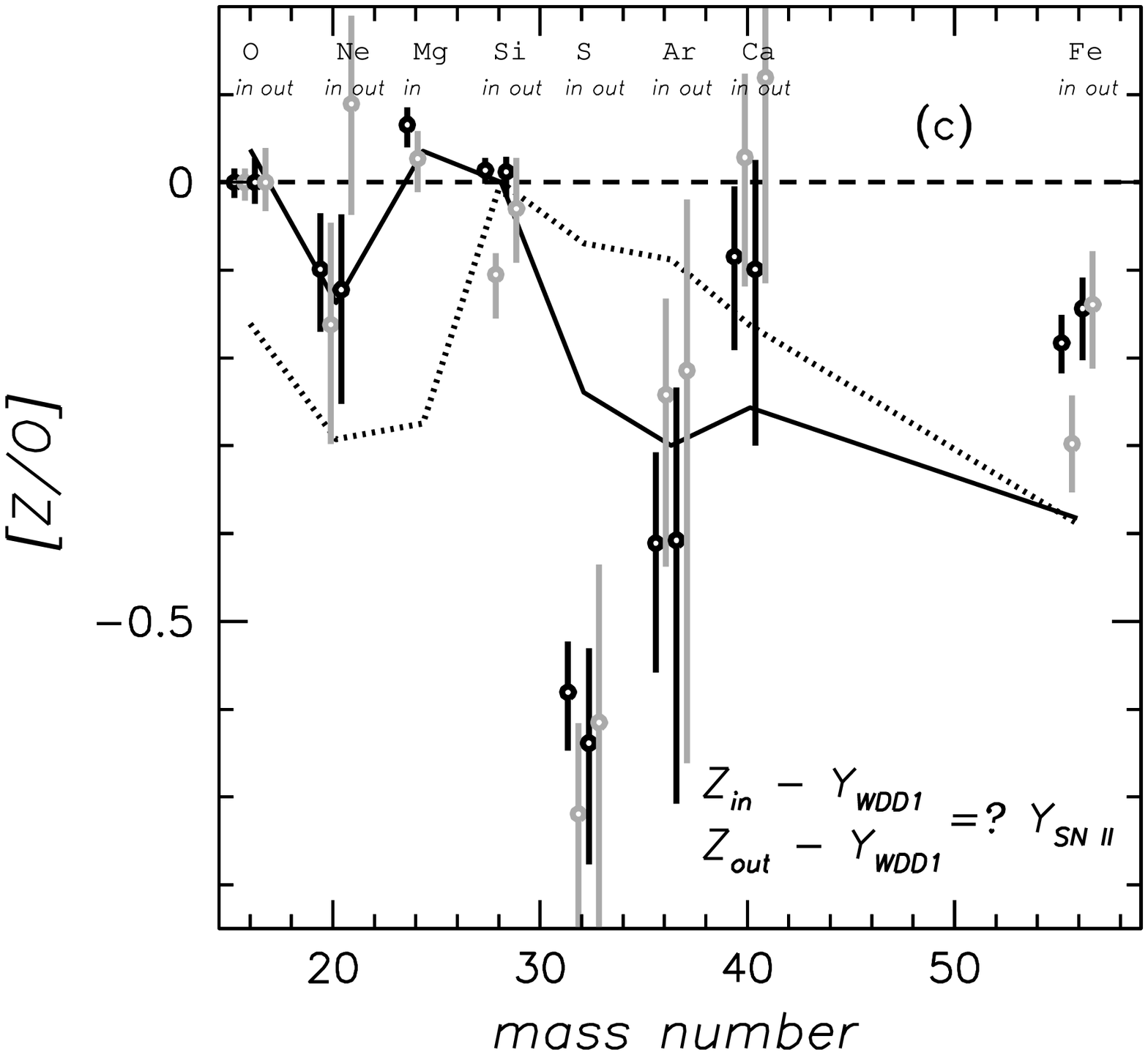}\hfill\includegraphics[width=7.cm]{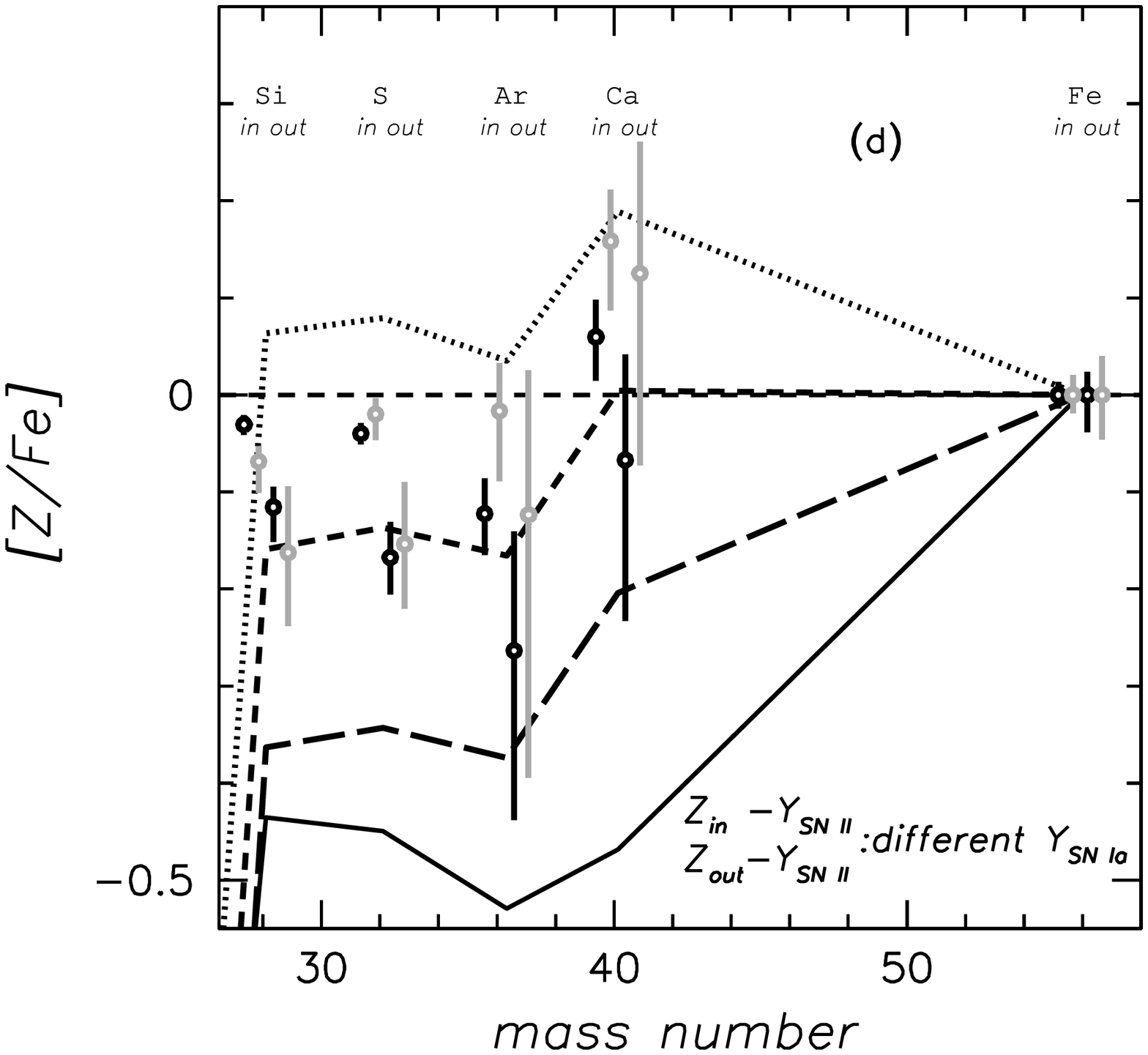}\hfill\hfill
 
 \hfill\includegraphics[width=7.cm]{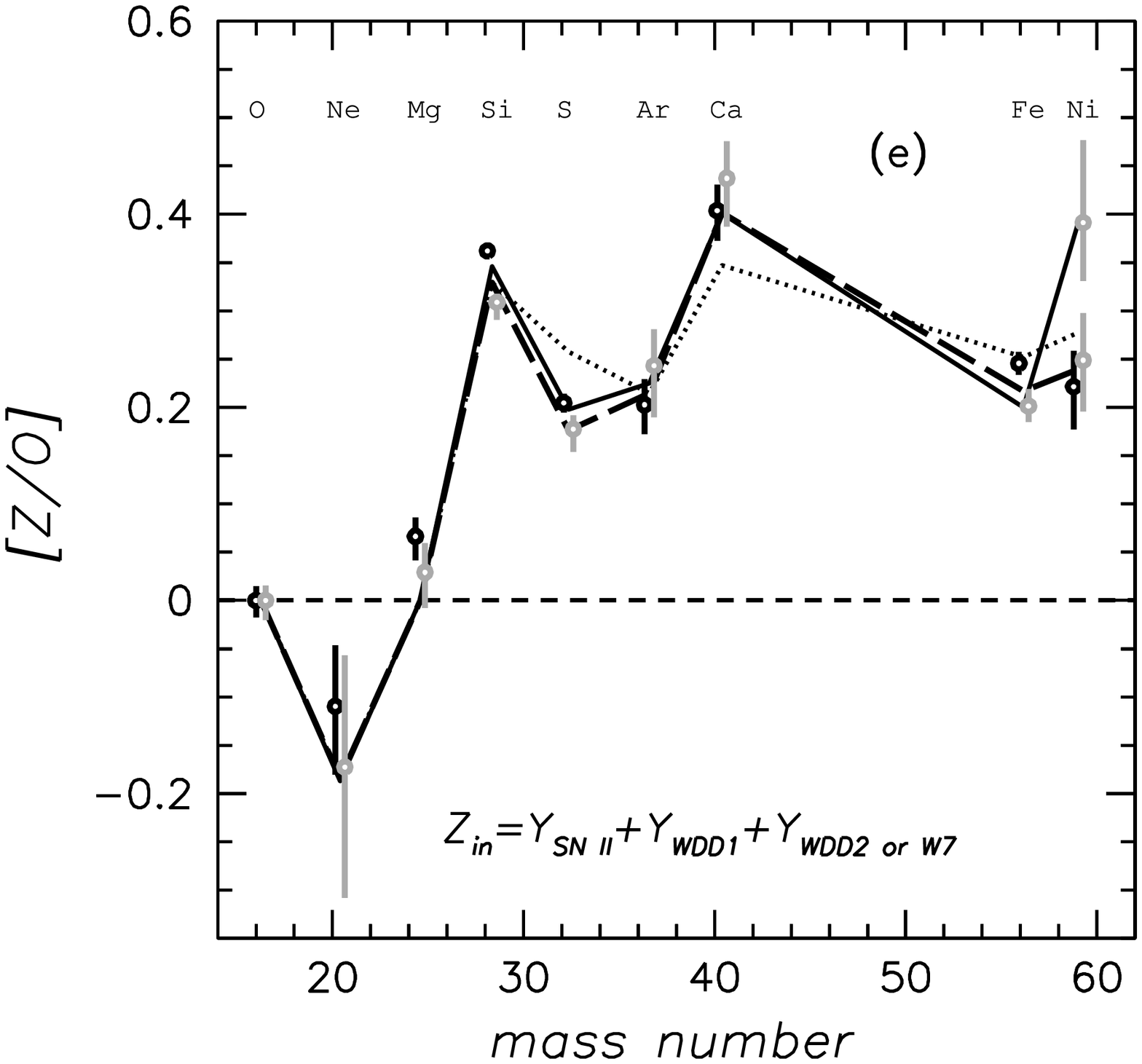}\hfill\includegraphics[width=7.cm]{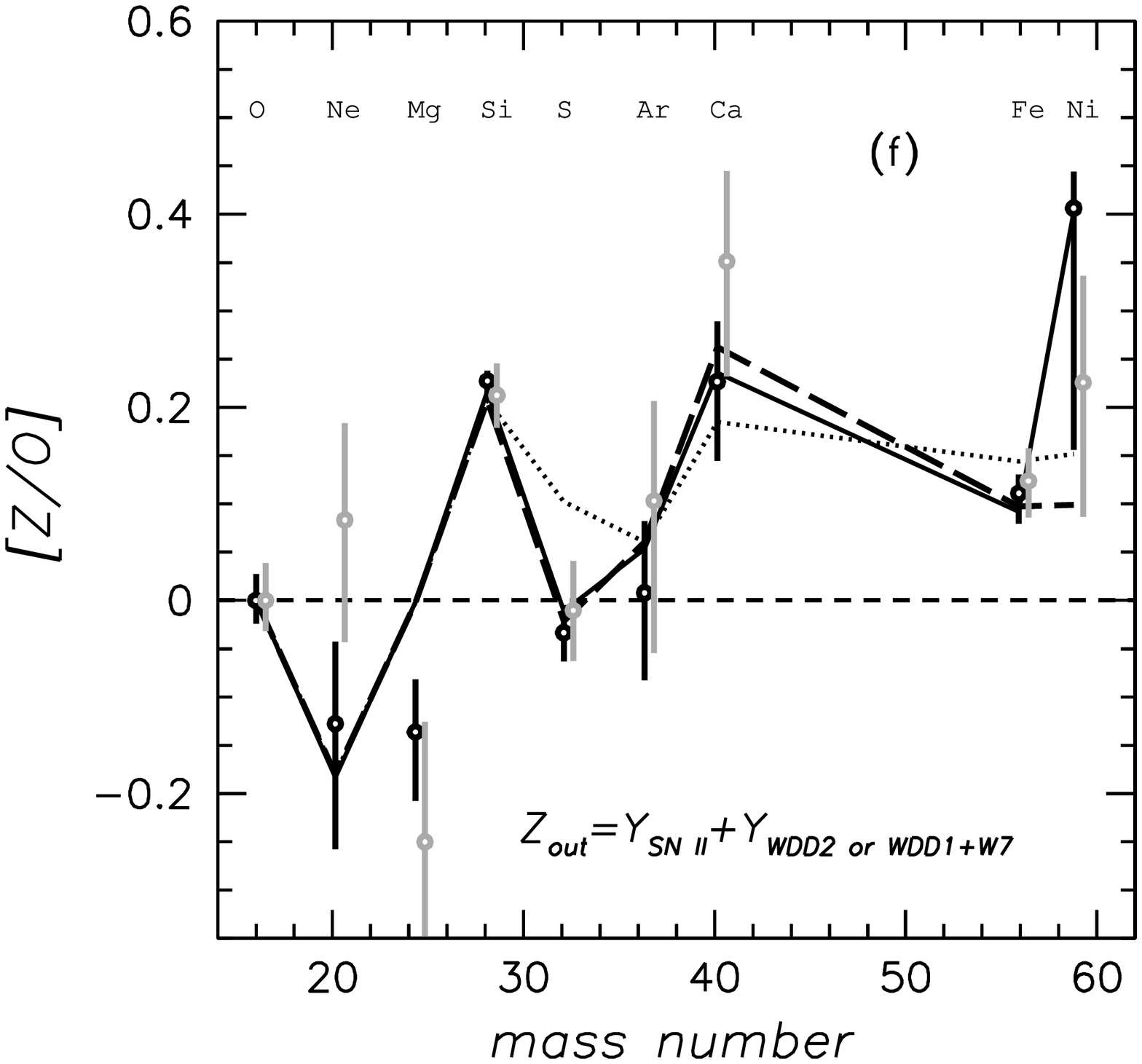}\hfill\hfill
%
%

\vspace*{-0.2cm}

\figcaption{A study of the SN composition of the observed abundances.
   Labeling at the bottom of each panel indicates the data and the modeling
   used, where $Z_{in}$ is the measurement in the central and $Z_{out}$ in
   the outer zones and $Y$ denotes SN yields. The line coding is similar for
   panels {\it (a,b,d)} and indicates W7 (solid), WDD1 (dotted), WDD2
   (dashed), WDD3 (long-dashed) SN Ia models of Nomoto \etal (1997). In panel
   {\it (c)} SN II yields are shown by the solid line (Nomoto \etal 1997a) and
   dotted line (Woosley \& Weaver 1995). Panels {\it (e)} and {\it (f)} show,
   respectively, the fits to $Z_{in}$ and $Z_{out}$, using modified SN II
   yields as discussed in the text and illustrating the change between
   combinations of WDD1 with WDD2 (long-dashed lines, dotted lines indicate
   Nomoto SN II yields) compared to WDD1 with W7 (solid lines) models. SNe
   number ratios for WDD1:W7:II are 10:2:20 and 5:3:20 and for WDD1:WDD2:II are
   8:3:20 and 0:7:20, for the inner and the outer zones, respectively.
\label{sn-all}
}
\end{figure*}

The structure of Tables \ref{tab:ab} and \ref{tab:ab2} is similar and is
defined as follows. In column 1 we identify the element, EPN measurements
are presented in columns 2 \& 3 and EMOS in columns 4 \& 5 for the central
and the outer zones, respectively.  Confidence intervals are stated at the
68\% level. As there is a difference between the definition of the solar
abundances used in the X-ray analysis and in theoretical modeling of
chemical enrichment (the first corresponds to photospheric values and second
to the meteoritic values from Anders \& Grevesse 1989), we provide in
columns 6 \& 7 the reference values for both, where for col 7 we use
tabulations of Woosley \& Weaver (1996). Major differences in the abundance
scales, are in iron (theoretical units are 1.45 times lower) and sulphur
(theoretical units are 1.14 times higher). While the data in columns 2--5
are given using the reference values in column 6, we will use throughout
this paper the theoretical scaling and also conventional units of elemental
abundance, $log(Z/Z_{\odot})$, marked with [] ([Z/H] in Fig.\ref{ab-data}
and [Z/Fe] and [Z/O] in Fig.\ref{sn-all}).

\section{Modeling}

\subsection{Test of SN Ia and SN II model yields}

In modeling the results, we aimed to test the SN Ia and SN II nucleosynthesis
models.  We illustrate our thread of thought in Fig.\ref{sn-all}, where the
labeling at the bottom right corner of each panel indicates the data and
modeling used. The label $Z_{in}$ indicates the measurement in the central
(inner) and $Z_{out}$ -- outer zones, and $Y$ denotes usage of theoretical
yields for SN II and different models of SN Ia of Nomoto \etal (1997) with
the differences outlined above.  Error bars are plotted at the 68\%
confidence level.

There is an increased role of SN II in the outer zone as compared to the
inner zone, as is best seen in the increase of [O/Fe] by almost 0.2 dex in
Fig.\ref{ab-data}.  To test the SN Ia yields, we first consider the inner
zone, after subtracting the SN II contribution to the observed abundance
patterns.  For that we have to i) choose the SN II yields and ii) to
normalize the SN II element patterns.  For the latter point we can use the
elements whose abundance is not affected by SN Ia enrichment.  Given the
possible systematic uncertainties in our measurements of Ne and Mg, we
choose O as such an index.  The modeling of SN II is quite complicated and at
present differs significantly among various authors (Gibson, Loewenstein,
Mushotzky 1997).  We thus used two methods to estimate the SN II contribution
and check the consistency of the results.

The first method (method 1) is independent of any modeling of the SN II
yields.  We subtract the outer zone element pattern, normalized to the
relative O abundance in the inner and outer zone.  The resulting element
pattern is shown in Fig.\ref{sn-all}a.  With this method we correctly
subtract the SN II contribution.  We note that indeed Ne and Mg (taking into
account that Mg/O ratio in the outer zone is consistent with the solar
value, as discussed in \S\ref{s:res}), which are enriched by SN II, are
consistently removed in this way. The drawback of this method is that we
also remove a fraction of the SN Ia contribution, existing in the outer
region.  The resulting patterns would be representative of the SN Ia
pattern, only if the SN Ia overall yields are the same in the inner and
outer regions.  They can very well differ, if the various SN Ia types
contribute in different proportions in the two regions.

The element abundance pattern, obtained in this way exhibits Si:S:Ar:Ca
proportions typical of SNe Ia yields, with a high Si/Fe ratio (0.1 dex),
corresponding well to WDD1 models of Nomoto \etal (1997). The high degree of
incompleteness of Si burning in SNe Ia, thus found for M87 center, is in
agreement with the conclusion based on the analysis of HST observations,
that SNe Ia in early-type galaxies are, in general, subluminous (Howell
2001).  In addition, another supporting argument comes from the
consideration of the Ni abundance in M87.  Although, we cannot produce a
meaningfully constrained difference in Ni abundance between the central and
the outer zone, the Ni/Fe ratio for the central zone, using EPN data on Ni
K-shell lines, is $1.5\pm0.3$ of the solar value.  Here errors are given at
the 68\% confidence. Abundance measurements are from K-shell lines for both
Ni and Fe and theoretical units for solar element abundance are used.  This
value is much lower than the Ni/Fe ratio of 3 times solar found by Dupke \&
White (2000a), which favors the W7 model of SN Ia by Nomoto \etal (1997).

The second method (method 2) depends on the SN II modeling, but is free of
any assumption on the radial variation of the SN Ia yields.  We subtract,
from the inner zone pattern, the SNe II yields from Nomoto \etal (1997a),
again normalized to the O abundance.  This is shown in Fig.\ref{sn-all}b.
This leads to a roughly similar abundance pattern as found with the previous
method, but significant differences are observed.  Let us first assume, for
the sake of the discussion, that the abundance pattern determined with the
first method gives correctly the SN Ia yields in the inner region, since a
good agreement is observed with the WDD1 model.  The differences observed,
when using method 2, would thus indicate incorrect SN II yield modeling.  By
comparing Fig.\ref{sn-all}a and Fig.\ref{sn-all}b, we first note that S is
over-subtracted.  Ca is under-subtracted relative to Si and Ar, although
marginally since the Ca yield in SN Ia based on the measurements in
Fig.\ref{sn-all}a has a large error bar.  Finally the Si/Fe ratio is by 0.1
dex lower, compared to Fig.\ref{sn-all}a, although only on the 90\%
confidence level, given large error bars on Fe in Fig.\ref{sn-all}a.  We now
examine if such discrepancies can be removed by modifying the corresponding
SN II yield for these elements in a reasonable way.

For sulphur, we require a reduction in SN II yields of Nomoto \etal (1997a)
by a factor of two to three in order to match the observations.  This,
indeed, has been known to be a problem of SN II modeling already from ASCA
observations of cluster outskirts, where the abundance pattern favors SN II
(the number ratio is 20:1 for SN II:Ia compared to 20:12 for the central
zone in M87) (Mushotzky \etal 1996; Loewenstein \& Mushotzky 1996; FDP).
For example from the Si:S:Fe ratios in the Centaurus cluster (Finoguenov,
Arnaud, David 2001) we conclude that Nomoto \etal (1997a) SN II yields for S
should be reduced by a factor of 2.5.  The resulting [S/Si] yield in SN II
is -0.6 dex.  Note that this makes S a good SN Ia indicator, especially in
the central zone of M87, where the S/Fe ratio by SN Ia takes a solar value.

We then consider the problem with the Si/Fe ratio apparent in
Fig.\ref{sn-all}b.  We again assume that the SN Ia yields are those from the
WDD1 SN Ia model and we illustrate the corresponding requirements on SN II
model yields, by subtracting the WDD1 SN Ia yields from the measured
abundance pattern in both the central and the outer zones.  The model
normalization is chosen to achieve the [S/Si] of -0.6 dex, as discussed
above.  The results are presented in Fig.\ref{sn-all}c.  We show the SN II
model yields of Nomoto \etal (1997a) with the solid line and the SN II
yields of Woosley \& Weaver (1995) for solar metallicity and averaged using
the Salpeter IMF with the dotted line.  Models are normalized to Si, since
SN II yields for Si are similar among different authors.  As is seen from
the Figure, to explain the data while assuming the WDD1 model for SN Ia
yields, the SN II plot should exhibit a high Fe/Si ratio. This is
unacceptable, given ASCA observations (FDP), as well as observations of the
stellar abundance pattern (\eg\ Timmes, Woosley and Weaver 1995). This
indicates that simply changing the SN II yields is not sufficient.  The
inconsistency between the abundance pattern derived with method 1 and method
2 is also due to a variation of the SN Ia yields with radius.

On the other hand, varying SN Ia models will not strongly help in reducing
the Ca/Si ratio in Fig.\ref{sn-all}c, so our observation hints toward a
solar Ca/Si ratio in SN II models, in agreement with stellar data (\eg\
Timmes \etal 1995). The Ca/O ratio is least dependent on the adopted S/Si
ratio and is still close to solar even when the sulphur production in SN II
is set to zero.  We also note, that O/Si ratio, which is claimed to be
uncertain by a factor of 3 ($\sim 0.5$ dex, Gibson, Loewenstein, Mushotzky
1997), is in fact very well reproduced in Nomoto \etal (1997a) SN II models
(O/Si ratio is reproduced to better than 10\%).  \footnote{Now we can see
that systematic difference in Si abundance between EPN and EMOS measurements
for the center is likely to be a problem for EPN (probably due to
systematics in the background subtraction), as already at the outer zone,
EPN value is on the same level as EMOS data.}

In a last step, we thus consider the Nomoto \etal (1997a) SN II model
yields, modified as suggested by our analysis: Si:Ca ratio is solar, while
[Si/S] is 0.6 dex.  We subtracted these yields both in the inner and outer
region (both shown in Fig.\ref{sn-all}d) and check the consistency of the
derived abundance pattern with the SN Ia theoretical yields.  The SN Ia
yields for the outer zone correspond well to the WDD2 model of Nomoto \etal
(1997). For the inner regions they lie in between the WDD1 and WDD2
model. We note that, in both cases, the Si:S:Ar:Ca ratios are correct. The
actual difference in the SN Ia yields between the central and the outer zone
amounts to 0.1 dex for Si group elements and is detected on a confidence
level exceeding 99\% for Si and S, and marginally for Ar and Ca.

At this point, we have established that our data exhibit a variation in SN
Ia model yields between the central and the outer zones, relative ratios for
Si:S:Ar:Ca are correct in SN Ia models of Nomoto \etal (1997), Si:Ca ratios
in SN II models should be solar within 0.1 dex, while [Si/S] is 0.6 dex
(similar to ASCA conclusion on [Si/Fe] FDP), and that Nomoto \etal (1997a)
SN II model yields consistently describe O:Ne:Mg:Si:Ar within 0.05 dex.  Our
conclusion regarding the Si-Ca group could be formulated as follows: peaks
on Si and Ca relative to Ar and S (as seen in Fig.\ref{ab-data}) are due to
SN II, not SN Ia models and should become even stronger in the cluster
outskirts, where the SN II abundance pattern dominates.

Finally, it should also be noted, that our conclusion regarding SN Ia
yields, can be based on Ar, which, as a noble gas, cannot be locked into
dust. So we can rule out the possibility that the abundance pattern in M87
is heavily modified by condensation into dust and dust sputtering (Allen and
Fabian 1998; Aguirre \etal 2001).

\subsection{Contribution from different SN Ia types}

As expected from the above analysis, the observation of M87 could be
described by a set of delayed detonation models.  Using the theoretical SN
Ia yields and the modified SN II yields, SN number ratios for WDD1:WDD2:II of
8:3:20 and 0:7:20, for the inner and the outer zones, respectively, fit the
data well, as illustrated in Figs.\ref{sn-all}e and \ref{sn-all}f.  In these
figures we also illustrate the change due to improved SN II yields (compare
the dotted and the dashed lines).

However, this is not the only solution.  The factor of two higher Ni/Fe
ratio, reported for some {\it other} clusters of galaxies by Dupke \& White
(2000a) using ASCA observations, requires a large variation of neutron rich
isotope production in SN Ia models, beyond the reach of delayed detonation
models in both the level of Ni/Fe found by Dupke \& White (2000a) and the
variation in Ni/Fe ratio by a factor of 2, implied by a comparison of M87
data with the data of Dupke \& White (2000a).  Therefore, to explain the
combined data, introduction of deflagration models of Nomoto \etal (1997) is
required.  The need for both deflagration and detonation scenarios in order
to explain the element abundance pattern independently confirms the
conclusion of Hatano \etal (2000) on the spectroscopic diversity of SN Ia.
Two scenarios of SN Ia explosion, indicated by their study and also supported
by X-ray observations, as summarized above, detonation and deflagration,
introduce two free parameters related to SN Ia modeling: the mixture of
delayed detonation models (controls the average Si/Fe groups abundance ratio
in delayed deflagration models) plus a relative role of
deflagration-to-detonation scenario (acts both on Si/Fe and Ni/Fe ratios).
So, for a final answer, both Si/Fe group ratios and Ni/Fe are equally
important.

To illustrate the point, in Figs.\ref{sn-all}f and \ref{sn-all}g we show
that the W7 model in combination with the WDD1 model can also describe the
observed data in the central and the outer zones.  SN number ratios for
WDD1:W7:II are 10:2:20 and 5:3:20 for the inner and the outer zones,
respectively.  It is clear that using the existing data on the Ni/Fe ratio
for the outer zone of M87 it is not possible to discriminate between WDD2
and a mix of WDD1 and W7 models of SN Ia, while the higher reliability of Ni
abundance determination using K-shell lines, detected by EPN at the M87
center, which results in Ni/Fe ratio of $1.5\pm0.3$ solar value is best
modeled by a mix with W7 model.

Another aspect, related to understanding of the role of SN Ia at larger
radii of clusters, is whether the SN Ia abundance pattern will continue to
change with radius, reaching the W7 model yields. Observationally, the SN Ia
contribution to Fe should fall less steeply compared to the S abundance.
ASCA data on the Centaurus cluster shows that these two fall similarly and
in fact the WDD2-like SN Ia abundance pattern holds out to $270h_{50}^{-1}$
kpc distance from the center. Data on MKW4 (FDP) are consistent with no
further change in SN Ia pattern, but reveal a slightly higher role of the W7
model in general.  An even higher role of the W7 scenario is suggested for
AWM7, as Si/Fe and S/Fe ratios attributed to SN Ia are by 0.1 dex lower than
the WDD2 model yields, and also in the AWM7 cluster center (Finoguenov \&
Ponman 1999).  Future XMM observations will therefore be of interest to
study another aspect related to WDD1:W7 mix -- a link to chemical evolution
of galaxies.

\subsection{Hypernovae}

The consideration of nova and sub-Chandrasekar mass models for Type Ia SNe
would be important in the explanation of the rare elements (Woosley \etal
1997), which is not relevant here. The metal production by hypernovae is not
considered here as these events are thought to be rare. If there is an
association of hypernovae with an early Population III, the level of
corresponding metal enrichment, assuming uniform mixing, should be a few
orders of magnitude lower than the level observed in M87 (Madau, Ferrara,
Rees 2000). Formal application of the hypernovae yields (as a substitution
for SN II), when normalized to O, result in overproduction of S and Ni and
are at odds with either Mg, Si or Fe. The most energetic cases of HN models
can be formally used in place of SN Ia WDD models, as they mostly produce
the Si--Ca group (Nakamura \etal 2000). Consideration of hypernovae in the
interpretation of element abundances determined in clusters of galaxies by
XMM is given by Loewenstein (2001). Given the essentially SN-II type origin
of hypernovae, it is logical to attribute their appearance in the ISM of M87
to the stellar mass loss, which implies that there is still (as we consider
a hypernova scenario unlikely) no extra energy input to ICM associated with
the observed level of chemical enrichment of the X-ray gas.

\section{Discussion}

\subsection{Implications}\label{sec:im}

The enhanced Si production in SN Ia, suggested by our observations, has a
strong influence on the determination of the role of SN Ia in element
production in clusters of galaxies, since Si and Fe abundances are the most
easily available for measurements at X-rays. Already at 70 kpc distance from
the center, the abundance pattern attributed to SN Ia is reminiscent of the
WDD2 yields. Therefore, in the analysis of the ICM abundance pattern, as
opposite to study of ISM of BCG, it is the WDD2 model yields that have to be
considered.

The low iron yield, corresponding to the WDD1 scenario ($\sim0.5$\msun\
compared to $\sim0.7$\msun\ in W7 model), helps to reduce a discrepancy
between the estimation of SN Ia rates and measured element abundances in
X-ray halos of early-type galaxies (for details see Arimoto \etal 1997;
Finoguenov \& Jones 2000 and \S\ref{sec:pop}). In view of the varying Fe
yield in SN Ia models, for SN Ia rate estimates it is better to use weighted
Si and Fe abundances, since $M_{Si}+M_{Fe}$ is constant in SN Ia models, as
almost no elements lighter than Si are produced in significant amounts, as
well as few elements heavier than Fe group, while the initial mass of the
exploding star is fixed. The weighting method is to use (after separation of
SN II contribution) half of the Si abundance, add Fe abundance and use an
effective yield for iron and silicon of 0.9 \msun.

Observational evidence for a high Ni/Fe ratio, present in the literature,
which is a specific feature of the W7 SN Ia model, could partly be due to
confusion with the effect of resonance line scattering. Our results on the
variation in SN Ia produced element abundance pattern implies that Ni/Fe can
vary by a factor. A proper justification, which requires either high spatial
(physical) resolution to study the abundance profiles in the central part of
cluster or high energy resolution to separate the contribution of the Fe
K$_{\beta}$ line from Ni K$_{\alpha}$ line, is needed to finally conclude on
both the SN Ia type and the effect of resonance line scattering.

Iwamoto \etal (1999) worked on the refining of the SN Ia models in
application to the chemical evolution of Milky Way.  There, SN Ia yields
span a narrower (by 0.2 dex) range in Si-Ca/Fe ratios, in comparison to
models of Nomoto \etal (1997), which is insufficient to describe our
measurements and therefore requires further investigation of the parameter
space in SN Ia models. We note for example that the lowest DDT density used
in their calculations is $1.7\times10^7$ g cm$^{-3}$. Thus, we conclude that
within the frame of modeling of Iwamoto \etal (1999), our data favor a
somewhat lower DDT density, which is in agreement with initial studies of
Khokhlov (1991). However, the exact relation between the DDT density and the
resulting Si/Fe ratio is strongly model dependent.

From the studies of Umeda \etal (1999) we conclude that the calibration of a
variety of SN Ia models against the solar abundance pattern is poorly
connected to the SN Ia abundance pattern observed in clusters of galaxies,
as there is a systematic variation in the channels leading to SN Ia
explosion. For example, the abundance pattern in M87 changing with radius
from prevalence of SN Ia to ''solar'' proportion of SN II and Ia and further
on to dominance of SN II, will at no point resemble the solar element
composition.

\subsection{Population synthesis view}\label{sec:pop}

Apart from the information we can gain about the importance of different SN
Ia and SN II models in the heavy element enrichment process in the
intergalactic medium, there are also some possible implications on the
stellar evolution scenario in M87 that can be drawn from the observed
abundance patterns. One aspect of the problem considered above essentially
consists of an explanation of the abundance pattern by enrichment with
mostly SN II and an additional contribution of SN Ia to adjust the deviation
in the abundance distribution.

We can now address some further questions related to the connection of the
intracluster gas and the stellar population and implications from a
comparison of absolute abundances between the stellar component and the
X-ray emitting gas. The current approach to the modeling of the X-ray
element abundance pattern consists of fitting the stellar abundance pattern
(thought to be dominated by SN II) with extra SN Ia added to the hot gas
(\eg\ Loewenstein \& Mathews 1991). As this approach fails in the attempt to
combine the X-ray abundance measurements, optical abundance measurements,
and results of SN Ia searches (Loewenstein \etal 1994; Arimoto \etal 1997;
Finoguenov \& Jones 2000), there is a need for modification in the
underlying assumptions, such as the SN rate -- iron production relation
(Finoguenov \& Jones 2000), metallicity dependence of SN Ia rate (Kobayashi
\etal 1998; Umeda \etal 1999; Finoguenov \& Jones 2000), closed-box
approximation of the X-ray halo (Davis \& White 1996), as well as complexity
of interpretation of the X-ray emission in early-type galaxies (Matsushita
\etal 2000), recently resolved with the advent of Chandra (\eg\ Finoguenov
\& Jones 2001), as well as a revision of the results form the optical
spectroscopy (Kobayashi \& Arimoto 1999). In this section we would like to
shed light on one more issue, concerning the relation of the observed
stellar metallicities to metallicity of the stellar mass loss.

Since M87 sits at the bottom of the Virgo cluster potential, we need to
justify that the X-ray gas at the center of M87 is related to the M87 ISM,
rather then the ICM of the Virgo cluster, to ensure the usage of the
closed-box approach. Matsushita (2001) analyzed a number of early-type
galaxies observed by ROSAT and established the direct correspondence between
the temperature of the galactic X-ray halo and the stellar velocity
dispersion (often expressed as $\beta_{spec}=1$). This component exhibits a
good correlation with the optical luminosity of the galaxy.  Thus, the
ability to explain the temperature of the central zone of M87 consistently,
is essential to ensure robustness of interpretation of the absolute level of
element abundances. In the center of M87, a cold X-ray component
($kT\sim0.8$ keV) is found by MBFB, which corresponds to
$\beta_{spec}=1$. This component is associated with the radio jet and thus
is protected from adiabatic compression and thermal conductivity, suggested
to act on the ISM of cluster galaxies (\eg\ Vikhlinin \etal
2001). Subsequent rise of the temperature could either be a result of
adiabatic compression or thermal conduction.  We note, however, that an
increased velocity dispersion is detected for globular clusters in outskirts
of M87 and the reconstructed behavior of the M/L ratio with radius matches
the X-ray determination (Romanowsky \& Kochanek 2001). So in fact the mass
loss from the stellar component corresponding to the same globulars should
result in the observed X-ray temperature without invoking any
cooling/heating into its explanation. SN Ia provide 0.4 and 0.2 keV/particle
at the central and outer zones, respectively, but this input may be spent on
adiabatic expansion work. Thus, the considerations given above are not
changed.  The observed element abundance could be diluted by the inflow of
the Virgo ICM. Within this scenario, we can use a systematic change in SN Ia
pattern with radius to determine the dilution factor. The idea is to
determine the contribution to Fe abundance from the SNe Ia pattern
characteristic to Virgo ICM in both the center and the outer zones. In the
center this value will be lower, because it is diluted by the stellar mass
loss, while additional enrichment from SN Ia has a different element
abundance pattern. We take the SN Ia element abundance pattern found in the
outer zone to represent the in-flowing gas and assuming that M87 SN Ia are of
WDD1 type. The dilution factor obtained in this way equals 2, and the
resulting stellar O abundance factor is 0.83 solar and SN Ia rate is $0.16$
SNu (supernova units), assuming $M_*/L_B=8$ (see Finoguenov \& Jones 2000
for details of the calculation). One can see that this modification,
although yielding a level for SN Ia rate, consistent with optical searches
(Cappellaro \etal 1997), still does not completely solve the discrepancy in
stellar metallicity.

Since the stellar abundance pattern is thought to be dominated by SN II and
given the significant input of SN Ia for the Si abundance, discussed above,
we select O for a comparison of the ISM and stellar abundances. We note that
the O abundance appears a factor of two lower than current optical estimates
of Kobayashi \& Arimoto (1999), assuming its slight oxygen overabundance
relative to iron in stars.

The deviation in the derived stellar abundance between the optical and X-ray
data is moderate and we note that an interesting solution is traced in M87
globulars. The latter display two major populations of metal rich (MRC) and
metal poor clusters (MPC). The former follows the distribution of the field
stellar population, while the latter is distributed differently than the
stellar light in M87 (Harris 2001). McLaughlin (1999) proposed to use the
X-ray gas in the estimation of the efficiency of the MPC production and
concluded that efficiency of globular cluster formation is universal. This
either means that the efficiency of star formation is not universal
(McLaughlin 1999), or, as we propose here, that the metal-poor field stellar
population, related to the MPC, had an IMF low-mass cut-off at high mass
(Larson IMF) and thus left a few stars shining until now. While the IMF
slope is governed by the turbulence pattern (\eg\ Elmegreen 1993), the
proposed bi-modality of the IMF finds its origin in the sensitivity of the
Jeans mass to the temperature of the medium. The formation of the GC
requires high density, so it is essentially insensitive to the temperature,
which makes GC better tracers of the star formation (Larson 1998). In the
proposed scenario, the metallicity of the X-ray gas would reflect the
abundance pattern of both metal-rich and metal-poor stellar populations,
while only the metal-rich field stellar population is optically seen. Since
populations of the MRC and MPC in M87 are equally abundant (Harris 2001), a
factor of two lower metallicity in X-rays could easily be explained. An
important aspect of this scenario is the accumulation time of the X-ray gas
in the central zone, since it should exceed the time since the putative die
out of the metal-poor field stellar population of M87. The influence of the
hidden stellar population on the chemical enrichment of the ICM and its
effect on the Faber-Jackson relation are discussed in Elbaz, Arnaud,
Vangioni-Flam (1995); Zepf \& Silk (1996); Larson (1998).

While the Larson IMF may not be the only explanation, metal-poor field stars
were likely accompanying metal-poor GC till a certain point, but may become
lost to the cluster potential (Elmegreen B. 2001, private communication).
This alternative does not change the consideration given above, as the X-ray
gas still represents the abundance of both metal rich and metal poor stellar
populations, while only the metal-rich field stellar population is observed
optically.

The possible introduction of hypernovae models in the explanation of the
abundance pattern in the central zone will not change the above
consideration, as they are not contributing to O, given the observed
metallicities for the Si--Ca group.

The scenario, presented above, suggests a difference in the element
abundance in the hot X-ray halos of galaxies with different properties of
GCs (a deviation between optical and X-ray abundance should correlate with
GC specific frequency), which is worth further investigation.

\section{Conclusions}

Taking advantage of a nearly constant temperature behavior in the $1'-3'$
and $8'-16'$ ($14'$ for EMOS) annuli in M87 (the central and the outer
zones), we have derived the detailed element abundance pattern in M87 using
XMM-Newton observations with EPIC-PN and EPIC-MOS detectors. Given the
prevalence of a SN Ia origin for most elements, we were able to put
model-independent constraints on the SN Ia yields, requiring, in particular,
a high Si, S, Ar and Ca to iron ratio in the central zone of M87.  We detect
a variation in the SN Ia yields between the central and the outer zones,
which we relate to an increased incompleteness of Si burning in the SN Ia,
responsible for the M87 ISM enrichment. This result is also related to
subluminous SN Ia found optically in early-type galaxies. In addition, the
Ni/Fe ratio, in the central zone of M87 is $1.5\pm0.3$ (meteoritic) solar,
while values around 3 times solar are reported for other clusters. In
modeling of SN Ia, this implies a reduced influence of fast deflagration SN
Ia models in chemical enrichment of M87's ISM.  Thus, to describe the
combined X-ray data on clusters of galaxies in terms of present SN Ia
nucleosynthesis models, both (fast) deflagration as well as delayed
detonation scenarios are required, independently supporting a similar
conclusion, derived from optical studies of SNe Ia. We propose to use the SN
Ia contribution to both Si and Fe abundance for X-ray estimates of the SN Ia
rate. Regarding the SN II yields, we demonstrate that the Si:Ca ratio should
be solar within 0.1 dex, while [Si/S] is 0.6 dex, similar to the ASCA
conclusion on [Si/Fe] (FDP), and that the Nomoto \etal (1997a) SN II model
yields consistently describe O:Ne:Mg:Si:Ar within 0.05 dex. Finally, we
compare the derived O abundance with metallicity measurements using
integrated optical light and speculate on the importance of the stellar mass
loss from the metal-poor stars that previously accompanied the metal-poor
globulars of M87.


\begin{acknowledgements}
  
  The authors thank Ken'ichi Nomoto, Chiaki Kobayashi, Nikolai Chugai, Bruce
  Elmegreen, Friedrich-Karl Thielemann, Paolo Mazzali and the referee,
  Michael Loewenstein, for their comments on the paper. The authors
  acknowledge useful communications of Elena Belsole regarding the issues of
  EMOS data analysis. The paper is based on observations obtained with
  XMM-Newton, an ESA science mission with instruments and contributions
  directly funded by ESA Member States and the USA (NASA). The XMM-Newton
  project is supported by the Bundesministerium fuer Bildung und Forschung,
  Deutschen Zentrum fuer Luft und Raumfahrt (BMBF/DLR), the Max-Plank
  Society and the Haidenhaim-Stiftung.  AF acknowledges receiving the
  Max-Plank-Gesellschaft Fellowship.

\end{acknowledgements}

\bibliographystyle{aabib99}

\end{document}